COMMUNICATION

# Integrated Bioelectronic Proton-Gated Logic Elements Utilizing Nanoscale Patterned Nafion

J.G. Gluschke[a], J. Seidl[a], R.W. Lyttleton[a, b], K. Nguyen[a], M. Lagier[a], F. Meyer[a], P. Krogstrup[c], J. Nygård[c], S. Lehmann[b], A.B. Mostert[d], P. Meredith[e,f] & A.P. Micolich[a,†]

A central endeavour in bioelectronics is the development of logic elements to transduce and process ionic to electronic signals. Motivated by this challenge, we report fully monolithic, nanoscale logic elements featuring *n*- and *p*-type nanowires as electronic channels that are proton-gated by electron-beam patterned Nafion. We demonstrate inverter circuits with state-of-the-art ion-to-electron transduction performance giving DC gain exceeding 5 and frequency response up to 2 kHz. A key innovation facilitating the logic integration is a new electron-beam process for patterning Nafion with linewidths down to 125 nm. This process delivers feature sizes compatible with low voltage, fast switching elements. This expands the scope for Nafion as a versatile patternable high-proton-conductivity element for bioelectronics and other applications requiring nanoengineered protonic membranes and electrodes.

**New concepts**

Nafion is an ionomer that is heavily used as a proton conductor in proton exchange membrane fuel cells due to its exceptionally high protonic conductivity. Nafion has recently generated considerable excitement as a potential component in artificial synapse devices. A major roadblock to these new applications is the lack of methods for scalable patterning of Nafion films into device structures and integrated circuits. Our new concept here is the demonstration of a scalable method for electron-beam processing of Nafion films spin-coated on solid substrates to give nanoscale patterns with linewidths as low as 125 nm. This new Nafion patterning capability enables its use as an ion-conducting element in complex nanoscale devices and integrated circuits in bioelectronics. We provide a proof-of-concept demonstration by producing a hybrid organic-inorganic ion-to-electron transducer circuit giving DC gain exceeding 5 and frequency response up to 2 kHz. This performance competes well with state-of-the-art ion-to-electron transducers featuring PEDOT:PSS. Our work highlights the outstanding potential of Nafion as a high-speed device-level ion-transport material for advanced bioelectronics applications.

## Introduction

A key challenge in the emerging discipline of bioelectronics is the connection and integration of read-write-process electronic circuitry directly or indirectly with a target biological entity.[1–3] The field sits at the nexus between the physical, chemical and life sciences, and holds considerable promise for the creation of disruptive technologies for healthcare.[4] Indeed, some claim that bioelectronics, and related endeavors such as bionics, are the next evolutionary phases of medicine.[4] A defining feature of bioelectronics is the diversity of charge carriers – in addition to the familiar electrons and holes that carry charge in solid-state electronic materials, a large cast of ions, e.g., $H^+$, $K^+$, $Na^+$, $Cl^–$, etc., are the signal carrying entities in biology.[1,5] Thus, a key challenge for the field is the efficient and effective interconversion between electronic and ionic signals, a process commonly referred to as 'transduction'. This challenge, although simply stated, is far from trivial since the underlying physics of electrons, being quantum mechanical in nature, is very different to the large, center of mass, classical behaviour exhibited by ions.

The requirements for practical bioelectronic applications are wide-ranging and include: biocompatibility, high electronic and ionic conductivity, long lifetime and stability under biological conditions, as well as respectable prospects for patterning, device fabrication and deployment on soft conformable substrates.[2,6,7] These requirements have driven the exploration and development of a vast array of soft, organic, ionically active materials. These are then incorporated into device architectures by considering complexities

[a.] School of Physics, University of New South Wales, Sydney NSW 2052, Australia.
[b.] NanoLund, Lund University, SE-221 00 Lund, Sweden.
[c.] Center for Quantum Devices and Station Q Copenhagen, University of Copenhagen, DK-2100 Copenhagen, Denmark.
[d.] Chemistry Department, Swansea University, Swansea SA2 8PP, Wales, U.K.
[e.] Physics Department, Swansea University, Swansea SA2 8PP, Wales, U.K.
[f.] School of Mathematics and Physics, University of Queensland, Brisbane, QLD 4072, Australia.
† Corresponding author: adam.micolich@nanoelectronics.physics.unsw.edu.au
Electronic Supplementary Information (ESI) available. See DOI: 10.1039/x0xx00000x

such as capacitive versus Faradaic effects,[8] and device operating concepts, a classic example being the choice between organic electrochemical transistors (OECT) and organic field-effect transistors (OFET).[9] Of the various bioelectronic materials, PEDOT-PSS is the most developed owing to its long history as an electronic material[10] and recent development towards neural interfacing applications.[1,11] However, neural interfacing is only one part of a greater bioelectronic mission that includes the development of protonic transistors,[12,13] ionic logic[14] and 'iontronic' devices,[15] biosymbiotic circuitry,[16] e.g., 'electronic plants',[17] synapse-like devices,[18,19] pH sensor and control systems,[20–22] and biologically-tunable light-gated transistors.[23]

Advances in electronics are not only driven by new materials and new developments in existing materials, but also by new materials combinations. An example is a recent shift from all-organic or all-inorganic systems towards a more hybrid organic-inorganic systems, where some identified deficiencies of organics, e.g., speed/solubility, are resolved using traditional inorganic semiconductors such as Si or III-Vs[3,21,23,24] or deficiencies of inorganics, e.g., slow ion diffusion, are resolved using organic ionomers such as Nafion.[19] For example, OECTs have high transconductances[25,26] but can lose their performance advantage to OFETs at higher frequencies.[9] The carrier mobility of organic semiconductors is also several orders of magnitude lower than for their inorganic counterparts, further limiting electronic performance, e.g., switching speed. Organic semiconductors are predominantly $p$-type and the relative lack of high performance $n$-type materials makes all-organic complementary circuits suitable for bioelectronics applications rare.[27] In this regard we note that complementary logic would be the most likely platform for integrated bioelectronic circuitry, and so the scarcity of $n$-type organics is problematic and an area of considerable activity in the synthetic chemistry community.[28] Sub-micron patterning can also be difficult for many organics due to incompatibilities with the solvents used in lithographic processing, thereby limiting ultimate device miniaturization and integration. That said, traditional inorganic semiconductors on their own are also unsuited to bioelectronics, showing poor ion permeability, problems with surface effects and dangling bonds, and insufficient mechanical flexibility – let alone potential cytotoxicity if intimately exposed at an interface.[7]

Here we report a hybrid organic-inorganic system that draws a series of 'best in category' materials together into a single, monolithically integrated, nanoscale-patterned ion-to-electron transduction platform with strong prospects for bioelectronic applications. For our ion-transport material we chose the synthetic polymer Nafion, which has an exceptionally high protonic conductivity (78 mS/cm),[29] more than an order of magnitude higher than other biopolymers.[30] Nafion has previously been used in artificial synapse devices.[18,19] These macroscopic devices were made by sandwiching either a 0.18 mm thick Nafion-117 membrane[19] or cellulose tissue wipe soaked in Nafion solution[18] between the two halves of the device or spin-coating a Nafion film over a prepatterned epoxy (SU-8) mask.[18] Although effective for proof-of-concept, it highlights a major challenge for Nafion as a device material – scalability and integration. We developed a process whereby Nafion can be used as a negative electron-beam lithography resist to enable patterned structures with lateral dimensions as small as 125 nm to address this challenge. We show that Nafion has sufficient solvent-resistance to tolerate subsequent lithographic steps, giving a major versatility advantage compared to many other ion-transport materials used in bioelectronics. For our electronic componentry, we use a III-V nanowire complementary circuit platform[24] developed such that the entire Nafion-plus-III-V architecture is monolithically integrated onto a single chip. We demonstrate an inverter circuit with DC gain exceeding 5 and AC frequency response up to 2 kHz for sine wave signals and hundreds of Hz for square wave signals. This frequency response is sufficient for neural sensing applications, where the oscillation spectrum for brain activity ranges from 1-200 Hz.[11] Response beyond 200 Hz is desirable, however, as it affords better signal fidelity. This has driven a race to produce high-sensitivity ion-to-electron transducers that operate up into the kHz, a feat that has only very recently been achieved with PEDOT:PSS OECTs.[31]

An important aspect to note on the novelty of this work is that we use Nafion itself as the electron-beam lithography resist to leave residual nanoscale Nafion structures that are subsequently used as ionically active elements for device applications. The Nafion acts as a negative tone resist, opposite to the more conventionally used positive resist polymethylmethacrylate (PMMA) used in electron-beam lithography. The negative tone means we only need to scan the regions we wish to retain such that scalability depends only on the total area of the Nafion features we wish to keep not the total substrate area. Our work is also very different, in both method and intended application, to previous work where freestanding commercial Nafion-117 membranes were patterned with nanoscale holes using either conventional PMMA-based electron-beam lithography[32,33] or high-energy electron-beam damage[34] to influence ion permeability for proton exchange membrane (PEM) fuel cell applications.

Our work here represents a major step forward on prior-art for both Nafion in a device context[18,19] and hybrid organic-inorganic nanowire transducers.[24] It represents what we believe to be the first integrated logic circuit capable of processing ion and electron signals simultaneously and an interesting alternative to PEDOT:PSS OECTs for bioelectronics operational into the kHz frequency regime. Our results also highlight the potential for Nafion as a nanoscale patterned material for use as an ionically functional element in other bioelectronic device architectures, or indeed as a structured proton membrane or electrode for energy storage applications.

## Results and Discussion

### Direct electron-beam patterning of Nafion

Figures 1A-C illustrate the approach used for direct patterning of a Nafion thin-film by electron-beam exposure. The thin-film of Nafion was produced by spin-coating commercially available 5% Nafion-117 solution on a SiO$_2$-on-$n^+$-Si substrate (Fig. 1A). Select areas were exposed with a 5-10 keV electron beam to a dose of ~10 μC/cm$^2$ using a Raith150-Two electron-beam lithography system (Fig. 1B). This causes the electron-beam exposed Nafion

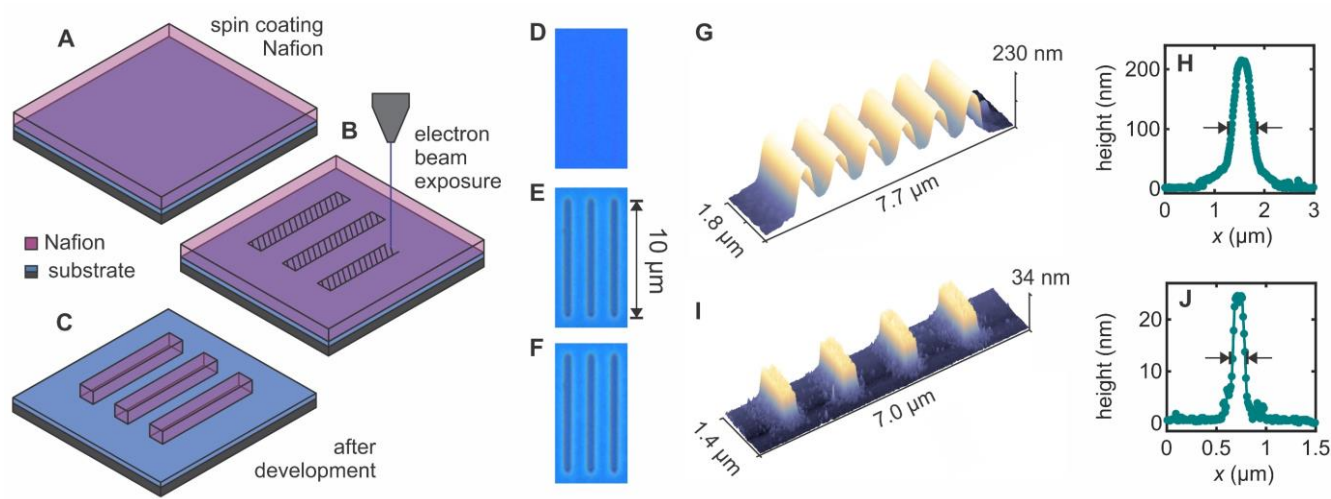

**Fig. 1** Nafion as a novel negative-tone electron-beam lithography resist (A-C) Illustration of direct electron beam patterning process. (A) Nafion is spin coated on the substrate. (B) Nafion channels are exposed with the electron beam (shaded area). (C) The unexposed nafion is dissolved in the 1:1 2-propanol:acetone developer leaving the exposed regions. (D-F) Optical microscopy images of a 230 nm thick Nafion film (D) immediately after electron-beam exposure, (E) after development, and (F) after prolonged exposure (see text) to the solvents 2-propanol, acetone, and $H_2O$ demonstrating the robustness of the electron-beam patterned Nafion. The substrate appears blue and the patterned Nafion purple. (G-J) Atomic Force Microscopy (AFM) scans and height profiles for (G/H) 230 nm thick and (I/J) 25 nm thick Nafion films. The AFM scans in (G) and (I) were obtained from lines with beam-patterned width 300 nm and pitch 1 μm for the 230 nm thick Nafion film and 1.5 μm for the 25 nm thick Nafion film. The height profiles (H) and (J) were obtained from a separate exposure where we wrote single lines with beam-patterned width 100 nm to determine ultimate resolution. The post-development full width at half maximum (FWHM) linewidths were: (G) 650 nm, (H) 420 nm, (I) 360 nm and (J) 125 nm.

regions to become comparatively insoluble, such that after brief immersion in a developer solution (1:1 2-propanol:acetone), the unexposed Nafion is dissolved away and only the exposed Nafion remains (Fig. 1C). In other words, we show that Nafion acts directly as a negative-tone electron-beam lithography resist.

Figures 1D-F demonstrate patterning of a 230 nm thick Nafion film produced by spin-coating undiluted 5% Nafion-117 solution at 3000 rpm for 30 s, followed by electron-beam exposure and development for 60 s in 1:1 acetone:2-propanol. The electron-beam-exposed region is evident by optical microscopy prior to development (Fig. 1D), due to a slight change in refractive index presumably arising from electron-beam induced material changes, and this region clearly remains undissolved after development (Fig. 1E). Notably, the electron-beam exposed regions are resistant to further solvent exposure thereafter. Figure 1F shows the same film after immersions in hot acetone (60°C) for 60 min, room-temperature 2-propanol for 60 min and room-temperature deionized water for 60 min, simulating the conditions typically found in the lift-off process for subsequent patterning of metal features. The Nafion features survive this treatment remarkably well, in stark contrast to other patternable ion-conducting polymers, e.g., polyethylene oxide.[24] This is a significant outcome because it means the Nafion deposition/patterning does not need to be the final process step, thereby making Nafion a far more versatile material from a fabrication perspective, as we will show below. The negative contrast of Nafion under electron-beam exposure is also fortuitous because it means that patterned regions can be written directly, making for a favourably scalable process compared to, for example, a positive contrast resist such as the polymethylmethacrylate (PMMA) typically used in electron-beam lithography.

On the issue of scaling, EBL systems are now routinely used in large-scale fabrication of, e.g., high-end photonic nanostructures and nanoimprint lithography masters,[35] providing substantial scaling potential for our Nafion patterning approach. Additionally, the direct negative-tone patterning of Nafion requires significantly fewer processing steps than indirect patterning via etch masks, e.g., as used for controlled roughening of commercial Nafion membrane surfaces.[32,33] This indirect patterning process is based on a positive tone EBL resist used to create a metal etch mask for oxygen plasma etching. This approach is not suitable for our devices because it would require EBL exposure of the entire wafer *except* for the nanoscale Nafion channels. Another surface roughening approach, developed for Nafion membrane and based on local destruction of material from high-energy electron beam exposure[34], shares this limitation. While these alternative approaches do not translate well to the device level, they underline a broader interest in nanoscale Nafion patterning.

Returning to the device fabrication, thinner Nafion films can be obtained by thinning the 5% Nafion-117 solution in ethanol,[36] and Figs. 1I/J demonstrate that the same electron-beam patterning process works well also for 25 nm Nafion films, giving improved linewidths. Figures 1G/I show Atomic Force Microscopy (AFM) scans obtained for line arrays with lithographic width 300 nm and 1 μm/1.5 μm pitch respectively for the 230/25 nm thick films. We obtain a full linewidth at half maximum of 650 nm for 230 nm-thick Nafion and 360 nm for 25 nm-thick Nafion for these arrays. Figures 1H/J show height profiles obtained for single lines with lithographic width 100 nm written to test ultimate resolution, giving a full linewidth at half maximum of 420 nm for 230 nm-thick Nafion and 125 nm for 25 nm-thick Nafion. Line broadening occurs in most EBL resists due to electron back-scattering from the substrate.[37] The 'teardrop'-shaped beam interaction volume[37] will naturally give broader lines for the 230 nm films (Fig. 1G). For adjacent lines to be ionically independent (i.e., the Nafion thickness reaches zero for a finite distance between lines) the separation presently needs to be approximately 300 nm and 1200 nm respectively for adjacent lines made in 25 nm and 230 nm thick Nafion films. We expect however that Nafion patterning resolution could be further improved by carefully optimizing exposure parameters such as beam energy, aperture, and dose, if required. Additionally, anisotropic etch methods, e.g., reactive ion etching, could be used to etch away the tails of the Gaussian line profile to ionically isolate adjacent lines at the base if finer line-separations are required. The devices presented in this work only use individual Nafion channels rather than arrays. Ionic crosstalk due to channel overlap is therefore not an issue presently.

The complete mechanism for the electron-beam produced solubility contrast is unclear at this stage but could be a combination of two contributions: a) direct chemical changes induced by the electron-beam, and b) structural changes due to local heating produced by electron-beam interactions. We explore these effects using x-ray photoelectron spectroscopy (XPS) and thermal annealing studies (see Supplementary Information).

For the XPS study, we compared electron-beam processed and unprocessed Nafion films. The data indicates cleavage of ether moieties on the sulfonated pendant groups. Nafion is a polymer featuring a polytetrafluoroethylene backbone with sulfonated perfluoropolyether pendant groups.[38] The insoluble fluorinated backbone is rendered soluble by the sulfonated pendant groups, meaning that pendant cleavage should produce a solubility contrast. Any associated crosslinking between Nafion moieties arising thereafter would add further to the contrast. We also note, the XPS data indicates that there is no change in the proportion of sulfonated groups to the rest of the polymer. This indicates that the proton concentration in the processed Nafion should remain unchanged after electron beam exposure.

Nafion is known to undergo structural changes that affect solubility[39] and protonic conductivity[29] by thermal annealing at temperatures below 100°C. Our thermal annealing tests on the solubility of unexposed spin-coated Nafion films in our developer solution and found that solubility is lost after a 10 minute bake at 100°C, but is retained even after several hours of baking at 60°C. It is difficult to measure the local heating generated during the electron-beam lithography process, but prior work suggests it is typically of order tens of degrees,[40] making our thermal test difficult to reconcile to our processing. For example, we note that our electron-beam energy and dose, at 10 kV and 10 $\mu C/cm^2$, are low by normal electron-beam lithography standards and our individual point exposure times are short (~3 $\mu s$). At this stage, while electron-beam induced chemical changes to the Nafion are the logical origin of our solubility contrast in light of conventional processes for electron-beam lithography and our XPS data, structural changes arising from local heating may also play some role.

More detailed studies of how electron-beam exposure affect the properties of our Nafion films will be reported separately.

**Managing Nafion acidity and the fabrication of monolithic nanowire proton-to-electron transducers**

The sulfonic acid groups in Nafion form hydrophilic pores that strongly adsorb water, to which the sulfonate acid groups donate $H^+$ ions, thereby providing Nafion with its exceptionally high protonic conductivity.[38] This presents an interesting twist from a device fabrication perspective, because it also means that Nafion is a solid superacid catalyst,[41] one that we found (unexpectedly) to rapidly dissolve InAs nanowires if either are connected to a large reservoir of electrons, i.e., any amount of metal (see Supplementary Fig. S8-S11). This process is essentially the equivalent of a metal-assisted chemical etch (MAC-etch),[42] but here implemented via a localized solid organic acid. This may have interesting prospects separate to this work, but for this project it was a challenge that we could only overcome because of the good solvent-resistance of the electron-beam processed Nafion described earlier. The Nafion needs to be patterned prior to any metal deposition if a nanowire is going to survive the fabrication process, which in turn means Nafion needs to be able to survive lithography processes. As we demonstrate below, the electron-beam processed Nafion can survive not just one, but three subsequent metallization steps in completing the device.

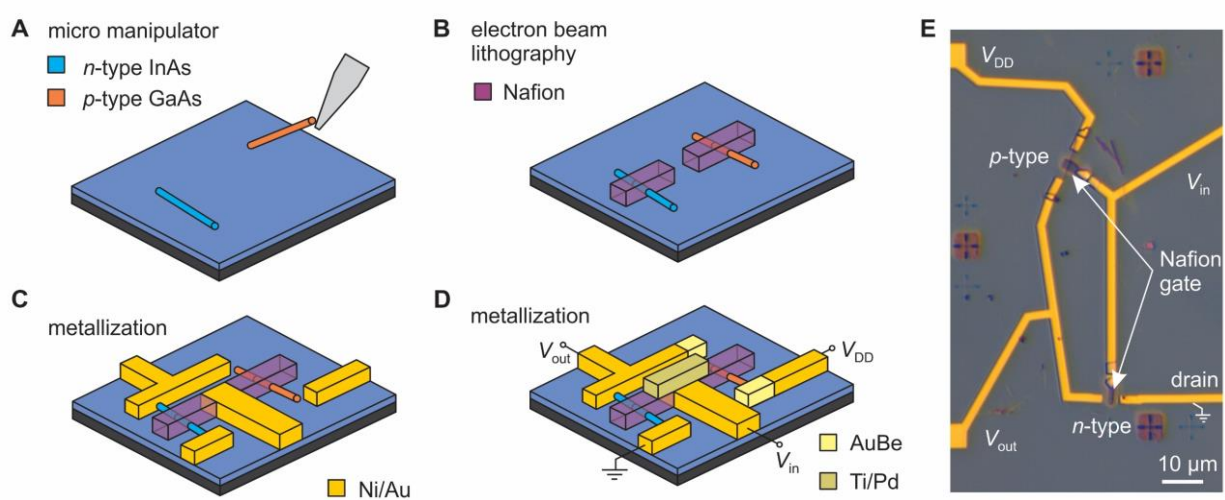

**Fig. 2** Process for making monolithic proton-to-electron transducers using Nafion and nanowires (A) InAs and GaAs nanowires were transferred to a pre-patterned $SiO_2$-on-n+-Si device substrate, (B) Nafion is spin-coated and electron-beam patterned, (C/D) three thermal evaporation steps follow to define (C) InAs ohmic contacts (Ni/Au), (D) GaAs ohmic contacts (AuBe) and interconnects and gate contact (Ti/Pd). Note that the Ti/Pd element can be written as Ni/Au as part of (C), as discussed in the text. (E) Optical micrograph of a completed inverter circuit with relevant features indicated.

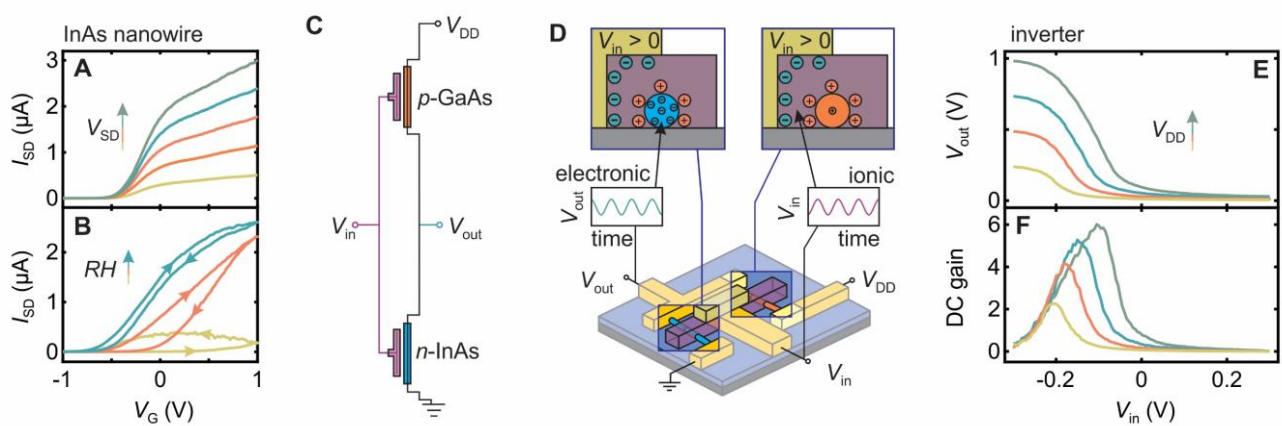

**Fig. 3** DC characterization of Nafion-gated InAs nanowire transistors and inverter circuits (A, B) Characterization of the current $I_{SD}$ vs voltage $V_G$ applied to the Nafion gate electrode at (A) five different source-drain biases $V_{SD}$ = 10, 20, 30, 40, 50 mV under maximum hydration conditions (relative humidity > 95%) and (B) at $V_{SD}$ = 50 mV for relative humidities RH = 50% (yellow), 70% (orange) and 90% (blue). (C) Inverter circuit diagram and (D) schematic of operation. An input signal $V_{in}$ modulates the electrical double-layers in the Nafion, which affects the nanowire conductivity producing an inverted copy of the input signal as the output $V_{out}$. (E) Output voltage $V_{out}$ vs input voltage $V_{in}$ and (F) DC gain $\partial V_{out}/\partial V_{in}$ vs $V_{in}$ under quasi-static (DC) operation for the inverter circuit at maximum hydration (RH > 95%). Data in (E, F) were obtained for four different drive voltages $V_{DD}$ = 250 mV (yellow), 500 mV (orange), 750 mV (blue) and 1 V (green).

Figures 2A-D show the fabrication process we ultimately used to produce fully monolithic proton-to-electron transducers using a two-nanowire complementary circuit (electronic side) gated by a nanoscale-patterned Nafion gate element (protonic side). We begin with a SiO$_2$-insulated $n^+$-Si substrate pre-patterned with large-scale interconnects and nanoscale alignment markers for accurate positioning. We use an micromanipulator[43] to place a small number of $n$-type InAs and $p$-type GaAs nanowires in each 100 × 100 μm² device field (Fig. 2A). We then register the nanowire position and pattern aligned Nafion strips using the procedure described earlier, prior to any metal contacts being added to the nanowires to prevent them being etched by the Nafion (Fig. 2B). Three rounds of electron-beam lithography and vacuum metal evaporation were used to: i) define Ni/Au ohmic contacts to the InAs nanowire and interconnects (Fig. 2C); ii) define AuBe ohmic contacts to the GaAs nanowire; and iii) define Ti/Pd contacts to the Nafion gate strips (Fig. 2D). The complete fabrication protocol is given in the Methods section. As an option, we can make the Nafion gate-strip contact out of Ni/Au and write it as part of the first lithography step (Fig. 2C). We have tried both approaches and find little functional difference thus far. An optical micrograph of a completed

functional device is shown in Fig. 2E. The active region has a typical footprint of 30 × 60 μm² (not including interconnects to external instrumentation). This could be further reduced by placing nanowires closer together using more accurate nanowire alignment procedures such as resist trench alignment.[44,45] Note that completed devices remain operational even after several months of storage in a N$_2$-glovebox and careful handling.

**DC electrical characterization of Nafion-gated nanowire transistors and inverter circuits**

Before examining the inverter circuit performance, we briefly demonstrate functionality of an individual Nafion-gated InAs nanowire transistor, since this a novel aspect of our work in itself. In Fig. 3A/B we plot the source-drain current $I_{SD}$ versus Nafion electrode voltage $V_G$ at different source-drain voltages $V_{SD}$ at maximal hydration (Fig. 3A) and different hydration levels for fixed source-drain voltage $V_{SD}$ (Fig. 3B) to highlight an important functional aspect of these devices. Proton conductivity in Nafion is facilitated by absorbed water, and is thus strongly humidity dependent.[29] All measurements were thus performed in a sealed chamber with gas inlet/outlet lines that enable the atmosphere to be vacuum purged, backfilled with nitrogen, and then hydrated by maintaining a continuous flow of nitrogen via a water bubbler. Under steady-state conditions, this gives a relative humidity exceeding 90%. As Fig. 3B demonstrates, gate action drops off sharply away from maximal hydration, indeed under ambient conditions our devices perform poorly if at all. This should not be viewed as a major problem – bioelectronics implicitly involves wet environments, if anything, our devices at maximum hydration are operated in a drier state than would be found in any bioelectronic application. The data in Fig. 3A shows that these devices perform well under maximal hydration conditions with sub-threshold slopes of typically 100-300 mV/dec and on-off ratios of $10^3$ and close to $10^4$ in our best devices. The operating range and performance is competitive with existing liquid electrolyte-gated nanowire transistors,[46–48] and excellent against equivalent all organic solid-state transducing transistors.[5,49] A more extensive characterization of a Nafion-gated nanowire field-effect transistor is provided in the Supplementary Information Figure S12.

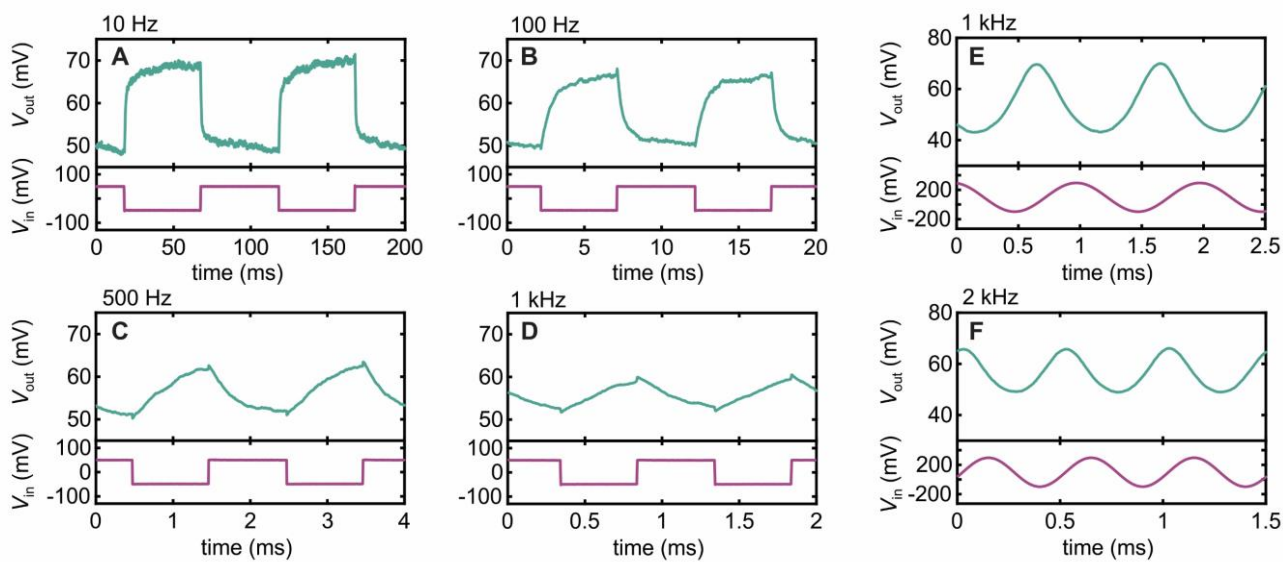

**Fig. 4** AC signal performance of Nafion-nanowire proton-to-electron transducer (A-D) Plots of $V_{out}$ (top) and $V_{in}$ (bottom) vs time at $V_{DD}$ = +1 V for square-wave input signals at (A) 10 Hz, (B) 100 Hz, (C) 500 Hz and (D) 1 kHz. (E/F) Plots of $V_{out}$ (top) and $V_{in}$ (bottom) vs time at $V_{DD}$ = +1 V for sine-wave input signals at (E) 1 kHz and (F) 2 kHz.

Our inverter is a simple integrated circuit consisting of a pair of nanowire transistors with opposite majority carrier-types, as shown in Fig. 3C, designed to take some input signal $V_{in}$ and invert the waveform, i.e., a maxima/minima in $V_{in}$ will give a minima/maxima in $V_{out}$. The action at a device level is illustrated in Fig. 3D; the mechanism is discussed in detail in the literature.[24,46,50] A potential difference across the Nafion strip drives electric double-layer formation at the electrode-Nafion and Nafion-nanowire interfaces. Note that although we use a metal electrode here, the potential difference could have an alternative origin, e.g., a neural action potential, as shown for OECT-based neural sensors.[11,26] The electric double-layer enacts a strong gating action on the nanowire, such that a positive (negative) voltage applied to the input electrode drives proton accumulation (depletion) at the Nafion-nanowire interface. This in turn drives electron accumulation (depletion) and enhanced (reduced) conductivity in the InAs nanowire,[46] and hole depletion (accumulation) and reduced (enhanced) conductivity in the GaAs nanowire.[47] From a circuit perspective, this drives $V_{out}$ towards ground ($V_{DD}$), such that when an AC input is supplied an inverted AC output signal results.

Turning to the complete inverter device, in Fig. 3E we plot the output voltage $V_{out}$ versus the input voltage $V_{in}$ for four different drive voltages $V_{DD}$. Sweeping $V_{in}$ between −0.3 V and +0.3 V is enough to modulate $V_{out}$ over 94% of $V_{DD}$, i.e., between 2% and 96% of $V_{DD}$. This indicates that the parasitic resistance in our inverter circuit is low, i.e., the dominant resistance contributions are from gateable channel material. In Fig. 3F we plot the DC gain $\partial V_{out}/\partial V_{in}$ versus $V_{in}$, with a 9-point smooth applied post differentiation to prevent point-to-point derivative noise from obscuring the signal. We obtain DC gain as high as 6, which compares well with the maximum gain recently obtained in all-organic liquid-electrolyte-based inverter circuits (~11)[27] and all-inorganic nanowire inverter circuits (~10).[51] The switching voltage ($V_{out} = V_{DD}/2$) occurs at $V_{in} < 0$ rather than the ideal $V_{in} = V_{DD}/2$, but this is not unusual for nanowire-based inverters.[51] It occurs due to an asymmetry in channel resistance between the $n$- and $p$-type nanowires,[51] an engineering problem resolvable by tuning the nanowire doping levels or nanowire length ratio to 'trim' the circuit into balance. Data from the individual InAs and GaAs nanowires in this device as well as DC characterizations of two additional inverter circuits is provided in Supplementary Information Figure S13.

**AC characterization of inverter performance**

Figure 4 shows the AC response of our inverter circuit at a range of frequencies obtained with $V_{DD}$ = +1 V. We begin with square-wave signals since those are the common test for inverters in the literature. We see respectable fidelity at low frequency $f$, with some slight rounding evident at 100 Hz due to capacitive effects. This exacerbates with increasing frequency, as expected, since the next edge occurs earlier in the exponential decay that arises from capacitive contributions as $f$ increases. Nonetheless, we can still distinguish a clear response to an input signal even at $f$ = 1 kHz, albeit with declining fidelity. Once we get above 1 kHz, parasitic capacitance contributions from the external circuit begin to contribute, which could be a significant part of the declining fidelity in Fig. 4A-D. At this point, one needs to remember that a square wave is a Fourier series of odd multiples of the fundamental, e.g., for a 1 kHz spectra, it contains significant components at 3 kHz, 5 kHz, 7 kHz, etc. As such, we reverted to sine wave signals in our examples at 1kHz and 2 kHz in Fig. 4E/F to better probe the ability of the Nafion to respond at a specific frequency. At both frequencies we obtain clean sine wave output, albeit with phase shifts of approximately 72° and 94° at 1 kHz and 2 kHz relative to low frequency operation, respectively, indicating a growing capacitive contribution to circuit impedance as the frequency is increased. However, in terms of the ability for Nafion to support charge alternation in the electric double-layers formed at its interfaces, the data in Fig. 4E/F indicates a response well into the kHz regime, a frequency range of interest for high-fidelity neural sensing applications,[11] that has only very recently been accessed using PEDOT:PSS OECTs.[31] Our work positions Nafion firmly as an alternative material for kHz-range bioelectronics applications.

## Conclusions

In this work we have demonstrated the nanoscale patterning of Nafion by utilizing it as a negative-tone electron-beam lithography resist. This enables direct-write of structures in a scalable manner, i.e., one only needs to write regions where Nafion should remain. After electron-beam exposure, the Nafion is remarkably resilient to subsequent fabrication processes including three further rounds of electron-beam lithography, vacuum thermal evaporation of metal, and lift-off in hot solvent. The Nafion also retains excellent protonic properties, and in our devices, the Nafion can respond well to sinusoidal AC signals at frequencies as high as 2 kHz. One challenge with Nafion, as we saw in the device fabrication, is managing its superacid catalyst properties. For our devices, this required careful ordering of the fabrication steps. Longer term, in completed devices, it might also present some issues that need to be managed. From a stability perspective, our devices remain functional after several months of storage under appropriate conditions, e.g., dry nitrogen, which prevents any attack of the Nafion on the nanowires and loss in contact performance, which is a widely known issue in nanowire devices. Finally, we note that the work here opens the path to devices beyond the Nafion-nanowire devices we have demonstrated. One could envision nanoscale artificial synapse devices similar to those reported at the macroscale by Josberger *et al*.[18] and van de Burgt *et al*.,[19] as well as potentially more complex bioelectronic devices incorporating protonic elements, e.g, protonic transistors.[12] There are also possibilities to combine nanoscale-patterned Nafion with ion-porous materials, e.g., PEDOT-PSS, assuming methods are found to pattern these at similar scales, to overcome challenges in the development of nanoscale bioelectronics. Such advances will be a focus of ongoing work.

## Materials and Methods

**Device Fabrication.** Device substrates were $n^+$-Si wafer (SVMI) coated on the front-side with 140 nm of thermal $SiO_2$ and on the backside with Ti/Au to enable the $n^+$-Si to be used as a global back-gate for devices if desired. The substrates were prepatterned with metal interconnects by photolithography and alignment markers by electron-beam lithography, as is standard for our nanowire devices.[52] Wurtzite-phase $n$-type InAs nanowires were grown by metal organic vapor phase epitaxy (MOVPE) using Au aerosols with a nominal diameter of 30 nm as seed particles.[53] An Aixtron 3 × 2 in$^2$ close-coupled shower-head system was operated at a total carrier gas flow of 8 slm and a total reactor pressure of 100 mbar. Growth commenced with a 10 min anneal in $AsH_3/H_2$ ambient at 550°C with an $AsH_3$ molar fraction of $\chi_V = 2.5 \times 10^{-3}$ before setting the temperature to 470°C for nanowire growth. A short InAs stem was grown at a $(CH_3)_3In$ molar fraction of $\chi_{III} = 1.8 \times 10^{-6}$ for 180 s with $\chi_V = 1.2 \times 10^{-4}$ before the wurtzite InAs nanowire was grown with $\chi_V = 2.3 \times 10^{-5}$ for 60 min. Growth was terminated by cutting the $(CH_3)_3In$ supply and cooling under $AsH_3/H_2$ ambient to 300°C. The GaAs nanowires were self-catalyzed and grown by molecular beam epitaxy on (111)Si.[54] The undoped core was grown at 630°C using $As_4$ and a V/III flux ratio of 60 for 30–45 min. The Be-doped shell was grown at 465°C using $As_2$ and a V/III ratio of 150 for 30 min giving nanowires with shell acceptor density $N_A = 1.5 \times 10^{19}$ cm$^{-3}$, diameter 120 ± 20 nm and length 5–7 μm. These nanowires are the highest doping density from earlier work on contacts to $p$-GaAs nanowires.[47,55] The nanowires are pure zinc blende crystal phase but may have short wurtzite segments at the ends; these wurtzite segments will be buried under the source/drain contacts. The InAs and GaAs nanowires were dry-transferred to the device substrate using a custom-built micromanipulator system,[43] which enables a small number of well-separated nanowires to be positioned with sufficient spatial control and then imaged, enabling us to keep track of $p$-GaAs versus $n$-InAs in subsequent processing. We used a 5% solution/suspension of Nafion-117 in lower aliphatic alcohols and water (a proprietary product by Sigma-Aldrich, 70160, Lot # BCBW9315). Nafion 117 has an equivalent weight of 1100 g/mol, which corresponds to 1100 grams of dry Nafion per mole of sulfonic acid groups when the material is in the acidic form.[38] The solution was used either as supplied or diluted up to 1:4 in ethanol to give thinner films. Undiluted Nafion solution spin-coated at 3000 rpm for 30 s with 1000 rpm/s ramp onto the substrates gives a 230 nm film, which is not baked prior to further use. Nafion film thickness was determined by ellipsometry using a J.A. Woollams M2000 spectroscopic ellipsometer and modelling the Nafion as a Cauchy material following Paul *et al*.[56] Electron-beam patterning of the Nafion is performed using a Raith150-Two electron-beam lithography (EBL) system with beam energy 5-10 keV, 15 μm aperture and dose 3-20 μC/cm$^2$ (typically <10 μC/cm$^2$). The device is then developed in a 1:1 mixture of acetone:2-propanol for 60 s at room temperature and dried. Three EBL-patterned metal depositions then follow to define: a) ohmic contacts to the InAs nanowires, b) ohmic contacts to the GaAs nanowires, and c) interconnects to the Nafion gate structure. The first metallization was vacuum thermal evaporation of 6 nm Ni and 134 nm Au performed immediately after a 60 s immersion in $(NH_4)_2S_x$ solution at 40°C to remove the InAs native oxide. The second metallization was vacuum thermal evaporation of 160 nm of AuBe alloy (1% Be in Au – ACI Alloys) performed immediately after a 30 s immersion in 10% HCl solution to remove the GaAs native oxide. The third metallization was 2 nm Ti and 60 nm Pd (if Ni/Au is used instead, it is deposited as part of the first metal deposition step) – the former was tested to promote absorption of hydrogen from the atmosphere to improve Nafion gating performance, however for this study, we found no significant performance gain from Ti/Pd over Ni/Au.

**Materials Characterization.** XPS was performed with an ESCALAB250Xi X-ray photoelectron spectrometer (Thermo

Scientific) with a mono-chromated Al K$\alpha$-source (energy 1486.68 eV). A beam with 500 µm spot size, 90° photoelectron take-off angle and 120 W power (13.8 kV × 8.7 mA) was used. Survey and high-resolution scans were obtained at 100 eV and 20 eV pass energies respectively. Data analysis of the XPS spectra was performed using Shirley background curves for the high-resolution scans. Peaks were modelled using pseudo-Voigtians with 30% Lorentzian character.

**Atmosphere control chamber.** All electrical measurements were conducted in an atmospheric control chamber. The chamber was built from vacuum fittings. Dry $N_2$ gas was run through a bubbler filled with deionized water and fed into the chamber inlet to increase the relative humidity (*RH*). An outlet on the other side of the system allows gas to escape to building exhaust. Humidity was monitored using a Sensirion SHT2x humidity sensor. The *RH* saturates at 90-96% after 30-60 min of $N_2$ flow. Unless otherwise specified, all electrical measurements were conducted under these conditions.

**DC single nanowire characterization.** Source-drain bias $V_{SD}$ was applied using a Keithley K2401 source-measure unit. The resulting drain current $I_D$ was measured with a Keithley 6517A electrometer. The gate voltage $V_G$ was applied using a Keithley K2400 source-measure unit enabling gate leakage current monitoring.

**DC inverter characterization.** Inverter drive bias $V_{DD}$ was applied using a Keithley K2401 source-measure unit. A DC input voltage $V_{in}$ was applied using a Keithley K2400 enabling gate leakage current monitoring, and the output voltage $V_{out}$ was measured using a Femto DLPVA voltage preamplifier and read out with a Keithley 2000 multimeter. The inverter drive was grounded via a Keithley 6517A electrometer to enable monitoring of the inverter drive current.

**AC inverter characterization:** Inverter drive bias $V_{DD}$ was applied using a Yokogawa GS200 source-measure unit. An AC input voltage $V_{in}$ was applied using a Stanford Research Systems DS345 signal generator, and the output signal $V_{out}$ was measured using a Femto DLPVA voltage preamplifier and read out using a National Instruments USB-6216 Data Acquisition Device with sampling rate 250 kHz. The input signal is also connected to the USB-6216 via a T-piece to enable simultaneous direct recording of both the applied input and resulting output.

## Acknowledgments

This work was funded by the Australian Research Council (ARC) under DP170104024 and DP170102552, the Welsh European Funding Office (European Regional Development Fund) through the Sêr Cymru II Program, the Danish National Research Foundation, the Danish Innovation Fund, NanoLund at Lund University, the Swedish Research Council, the Swedish Energy Agency (Grant No. 38331-1) and the Knut and Alice Wallenberg Foundation (KAW). P.M. is a Sêr Cymru Research Chair and an Honorary Professor at the University of Queensland and A.B.M. is a Sêr Cymru II fellow and the results incorporated in this work have received funding from the European Union's Horizon 2020 research and innovation program under the Marie Skłodowska-Curie grant agreement No 663830. A.P.M. was a Japan Society for the Promotion of Science (JSPS) Long-term Invitational Fellow during the drafting of this manuscript. The work was performed in part using the NSW and Queensland nodes of the Australian National Fabrication Facility (ANFF) and the Electron Microscope Unit (EMU) within the Mark Wainwright Analytical Centre (MWAC) at UNSW Sydney.

## Author contributions

A.P.M. and P.M. conceived and oversaw the research. J.G.G. contributed to all aspects of the experimental work, R.L. developed the initial patterning process for Nafion, J.S., K.N., M.L. and F.M contributed to various aspects of the fabrication and measurement of the nanowire-Nafion inverters. P.K., J.N. and S.L. contributed to growth of the nanowire materials, A.B.M. contributed to process development of the Nafion film, materials/electrical characterization and XPS analysis. All authors provided input during the manuscript preparation and revisions.

## Conflicts of interest

There are no conflicts to declare.

## Notes and references